\documentclass[aip,reprint,nofootinbib]{revtex4-2} 

\usepackage[utf8]{inputenc}
\usepackage[T1]{fontenc}
\usepackage{textcomp}
\usepackage{graphicx}
\usepackage{amsmath}
\usepackage{mathtools}
\usepackage{color}
\usepackage[colorlinks=True]{hyperref}

\draft 

\usepackage{etoolbox}

\makeatletter
\def\@email#1#2{%
 \endgroup
 \patchcmd{\titleblock@produce}
  {\frontmatter@RRAPformat}
  {\frontmatter@RRAPformat{\produce@RRAP{*#1\href{mailto:#2}{#2}}}\frontmatter@RRAPformat}
  {}{}
}%
\makeatother

\begin{document}


\title{Low intensity saturation of an ISB transition by a mid-IR quantum cascade laser} 



\author{Mathieu Jeannin}
\email[]{mathieu.jeannin@universite-paris-saclay.fr}
\author{Eduardo Cosentino}
\author{Stefano Pirotta}
\author{Mario Malerba}
\affiliation{Centre de Nanosciences et de Nanotechnologies, CNRS UMR 9001, Université Paris Saclay, 10 Boulevard Thomas Gobert, 91120 Palaiseau, France}

\author{Giorgio Biasiol}
\affiliation{Laboratorio TASC, CNR-IOM, Area Science Park, S.S. 14 km 163.5, Basovizza I-34149 Trieste, Italy}

\author{Jean-Michel Manceau}
\author{Raffaele Colombelli}
\email[]{raffaele.colombelli@u-psud.fr}
\affiliation{Centre de Nanosciences et de Nanotechnologies, CNRS UMR 9001, Université Paris Saclay, 10 Boulevard Thomas Gobert, 91120 Palaiseau, France}

\footnote{Authors to whom correspondence should be addressed: Mathieu Jeannin (\href{mailto:mathieu.jeannin@universite-paris-saclay.fr}{mathieu.jeannin@universite-paris-saclay.fr}) and Raffaele Colombelli (\href{mailto:raffaele.colombelli@u-psud.fr}{raffaele.colombelli@u-psud.fr})}

\date{\today}

\begin{abstract}
We demonstrate that absorption saturation of a mid-infrared intersubband transition can be engineered to occur at moderate light intensities of the order of 10-20 kW.cm$^{-2}$ and at room temperature. 
The structure consists of an array of metal-semiconductor-metal patches hosting a judiciously designed 253~nm thick GaAs/AlGaAs semiconductor heterostructure. At low incident intensity the structure operates in the strong light-matter coupling regime and exhibits two absorption peaks at wavelengths close to 8.9 $\mu$m. Saturation appears as a transition to the weak coupling regime - and therefore to a single-peaked absorption - when increasing the incident intensity. Comparison with a coupled mode theory model explains the data and permits to infer the relevant system parameters. 
When the pump laser is tuned at the cavity frequency, the reflectivity decreases with increasing incident intensity. When instead the laser is tuned at the polariton frequencies, the reflectivity non-linearly increases with increasing incident intensity. At those wavelengths the system therefore mimics the behavior of a saturable absorption mirror (SESAM) in the mid-IR range, a technology that is currently missing.
\end{abstract}


\maketitle 



Absorption saturation is one of the simplest non-linear optical process. 
In the case of interband transitions in semiconductors it relies on Pauli blocking of photoexcited carriers, 
reducing the ability of the material to absorb light. 
This saturation process is in general governed by the lifetimes and 
oscillator strength of the involved transition, and the frequency of the exciting light. 
Absorption saturation of interband transitions in semiconductors quantum wells (QWs) 
can be triggered with lasers at moderate intensities due to the long exciton lifetimes, 
which fill up the final states in the conduction band. 
Conversely, intersubband (ISB) transitions in doped QWs present very short (ps-scale) lifetimes and 
typical low photon energy of tens or hundreds of meV. 
Contrary to interband transitions, absorption saturation of ISB transitions 
does not stem from a filling of the upper state of the transition, but from a depletion of the available 
carriers in the ground state whose density is fixed by the doping introduced in the structure. 
These characteristics leads to high saturation intensities, in the sub-MW.cm$^{-2}$ range 
that are only attained with very high power lasers like 
CO$_2$ laser \cite{julien_optical_1988}, OPOs \cite{seilmeier_intersubband_1987, vodopyanov_intersubband_1997, zanotto_ultrafast_2012, mann_ultrafast_2021} or even free electron lasers \cite{helm_complete_1993}. 
These high intensity values are {\it in principle} not compatible with conventional semiconductor lasers output powers, 
which has so far limited the application of ISB-based saturable absorbers. 
Some recent work mention saturation-like behavior of ISB transitions triggered using a simple QCLs, 
but in that case saturation was considered a drawback for non-linear light generation \cite{lee_giant_2014, gomez-diaz_nonlinear_2015}.

ISB transitions are recognized for being a technological cornerstone in 
mid-infrared (mid-IR) optoelectronics, with the the invention of the quantum cascade laser \cite{faist_quantum_2018} 
that led to the development of practical mid-IR semiconductor lasers\cite{baranov_semiconductor_2013}. 
Since then, ISB transitions have contributed to the extension of semiconductor optoelectronic devices 
to the mid-IR and THz ranges of the electromagnetic spectrum. 
Examples are the demonstration and commercialization of mid-IR photodetectors\cite{schneider_quantum_2007, hakl_ultrafast_2021, hillbrand_femtosecond_2021, lagree_direct_2022, quinchard_high_2022}, 
free-space amplitude modulators \cite{pirotta_fast_2021, dely_10_2021}. 
As far as mid-IR saturable absorbers are concerned, initial results on narrow gap semiconductors \cite{vodopyanov_passive_1991} lead to recent efforts focused on finding new material systems 
relying e.g. on type II superlattices \cite{qin_semiconductor_2022} 
or two dimensional materials like graphene\cite{bao_atomiclayer_2009} and 
transition metal dichalcogenides\cite{tan_tuning_2017} (see Ref.\cite{ma_review_2019, liang_midinfrared_2021} for recent reviews). 
Hence, reducing the absorption saturation intensity in ISB devices is technologically appealing 
as it would allow to exploit the design flexibility of ISB transitions to implement mid-IR saturable absorbers. 
Demonstrating low-intensity saturation of an ISB-based devices is therefore an important first step.

In a recent theoretical work, we have developed a unified formalism to describe absorption 
saturation of ISB transitions coupled to microcavities\cite{jeannin_unified_2021}. 
In particular, we have shown that the nature of absorption saturation radically changes in cavity-coupled 
systems depending on the light-matter interaction regime, quantified by the vacuum Rabi frequency $\Omega_R$: 
\begin{equation}
\Omega_{R}^2 = f_w f_{12} \frac{\Delta n ~e^2}{4 \varepsilon \varepsilon_0 m^{\ast} L_{qw}} \equiv a \Delta n \label{eq:OmegaR}
\end{equation}
where $f_w$ is the filling fraction of the QWs inside the active region, $f_{12}$ is the transition oscillator strength hereafter approximated 
to the infinite square QW limit ($f_{12}=0.96$), $e$ and $m^{\ast}$ are the electron charge and 
effective mass, $\varepsilon_0$ and $\varepsilon$ are the vacuum permittivity and the semiconductor relative permittivity,
$L_{qw}$ is the QW width, and $\Delta n = n_1 - n_2$ is the population difference between the first ($n_1$) 
and the second ($n_2$) electronic levels of the QW. 
The absorption saturation manifests itself with the collapse 
of the light-matter coupling as both $\Delta n$, and thus $\Omega_R$, tend to zero.

In this work, we experimentally demonstrate that the absorption saturation of mid-IR 
intersubband transitions can be drastically reduced down to the kW.cm$^{-2}$ range. 
Such low values are compatible with the output power of commercially available quantum cascade lasers (QCL), 
interband cascade lasers (ICL) and emerging mid-IR fiber lasers\cite{woodward_modelocked_2018,matthew_infrared_2018}. 
The saturation intensity $I_{sat}$ is defined as the intensity for which the population difference $\Delta n$ is reduced to half the value of the introduced doping $n_s$, or - equivalently - the population in the excited level is $n_s/4$:
\begin{equation}
\Delta n_{I_{sat}} = \frac{n_s}{2} \iff n_{2,sat} = \frac{n_s}{4} .
\label{eq:saturation_condition}
\end{equation} 
To achieve saturation with a low incident intensity, a careful balance has to be found. 
Following Ref.\cite{jeannin_unified_2021}, structures with low doping can exhibit 
low saturation intensities when introduced in a properly designed cavity.
However, if the doping is so small that the polariton splitting cannot be resolved, 
the spectral signature of the saturation is masked by the cavity absorption. 
Increasing the doping allows to increase the frequency splitting, 
but leads to a linear increase in saturation intensity as $I_{sat} \propto \Omega_R^2$. 
The best choice is therefore to operate at the onset of the strong light-matter coupling regime, 
where a moderate $\Omega_R$ allows to spectrally resolve the polariton states while retaining a low $I_{sat}$. 
This constitutes a first steps towards a unique semiconductor saturable absorber mirror\cite{keller_solidstate_1992, keller_semiconductor_1996, keller_recent_2003} (SESAM) technology that could have great impact for mid-IR lasers.


%
\begin{figure}[!ht]
\centering
\includegraphics[width=\columnwidth]{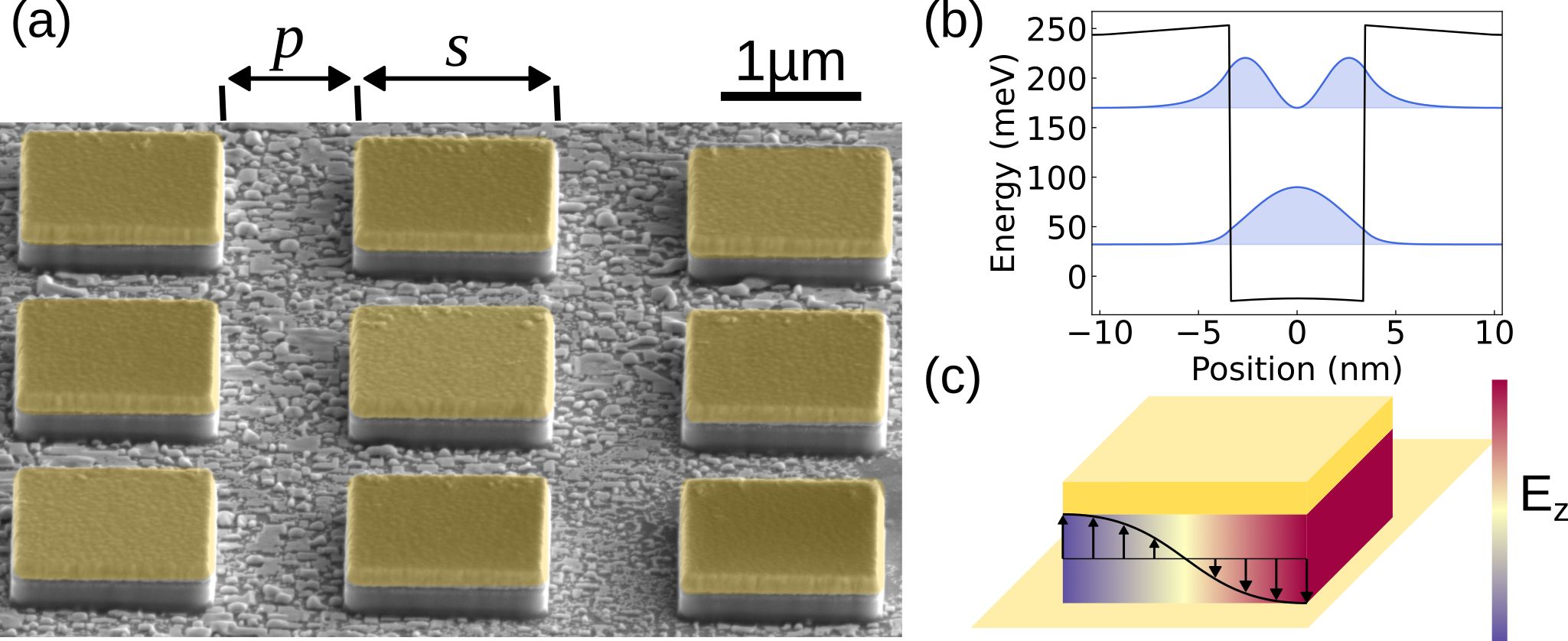}
\caption{\label{fig:intro_system}
(a)~Colorized SEM image of a typical patch antenna array defining the dimensions $s$ and $p$. 
(b)~Self-consistent Schrödinger-Poisson simulation of a period of the structure showing the two confined electronic states. 
(c)~Sketch of a single patch antenna with the electric field amplitude of the fundamental TM$_{00}$ mode. 
}
\end{figure}

We have designed a saturable absorber active region composed of seven repetitions of 
GaAs/Al$_{0.33}$Ga$_{0.67}$As QWs (6.5/14~nm, total thickness 253~nm) delta-doped with Si atoms 
with surface density of 2 10$^{11}$~cm$^{-2}$ (sample HM4448). 
Additional delta-doping layers on each side of the multi-QW structure are introduced 
to compensate for Fermi level pinning at the semiconductor-metal interfaces. 
A reference sample (HM4445) with the same, but undoped, active region has also been fabricated 
to measure the optical response of the bare cavity. 
The active region is processed using standard Au-Au thermo-compressive wafer bonding\cite{manceau_resonant_2018}, selective substrate removal in citric acid and electron beam lithography to fabricate arrays of 
metal-metal patch cavities as shown in Fig.~\ref{fig:intro_system}(a). 
The semiconductor between adjacent Au patches is removed by ICP-RIE using SiCl$_4$. 
A self-consistent Schrödinger-Poisson simulation (Fig.~\ref{fig:intro_system}(b)) 
shows the conduction band profile and confined electronic states of the QWs, 
with an energy separation of 138~meV corresponding to a transition wavenumber 
of 1110~cm$^{-1}$ ($\lambda \approx $ 9~{\textmu}m). 
The patch cavity size $s$ ranges from 1.25~{\textmu}m\ to 1.35~{\textmu}m\ to tune the frequency of the
fundamental TM$_{00}$ (Fig.~\ref{fig:intro_system}(c)) mode across the ISB transition. 
The distance $p$ between adjacent cavities is fixed at $p = 1$~{\textmu}m\ to operate close to the critical coupling condition, ensuring maximal energy funneling in the system.

\begin{figure}
\centering
\includegraphics[width=\columnwidth]{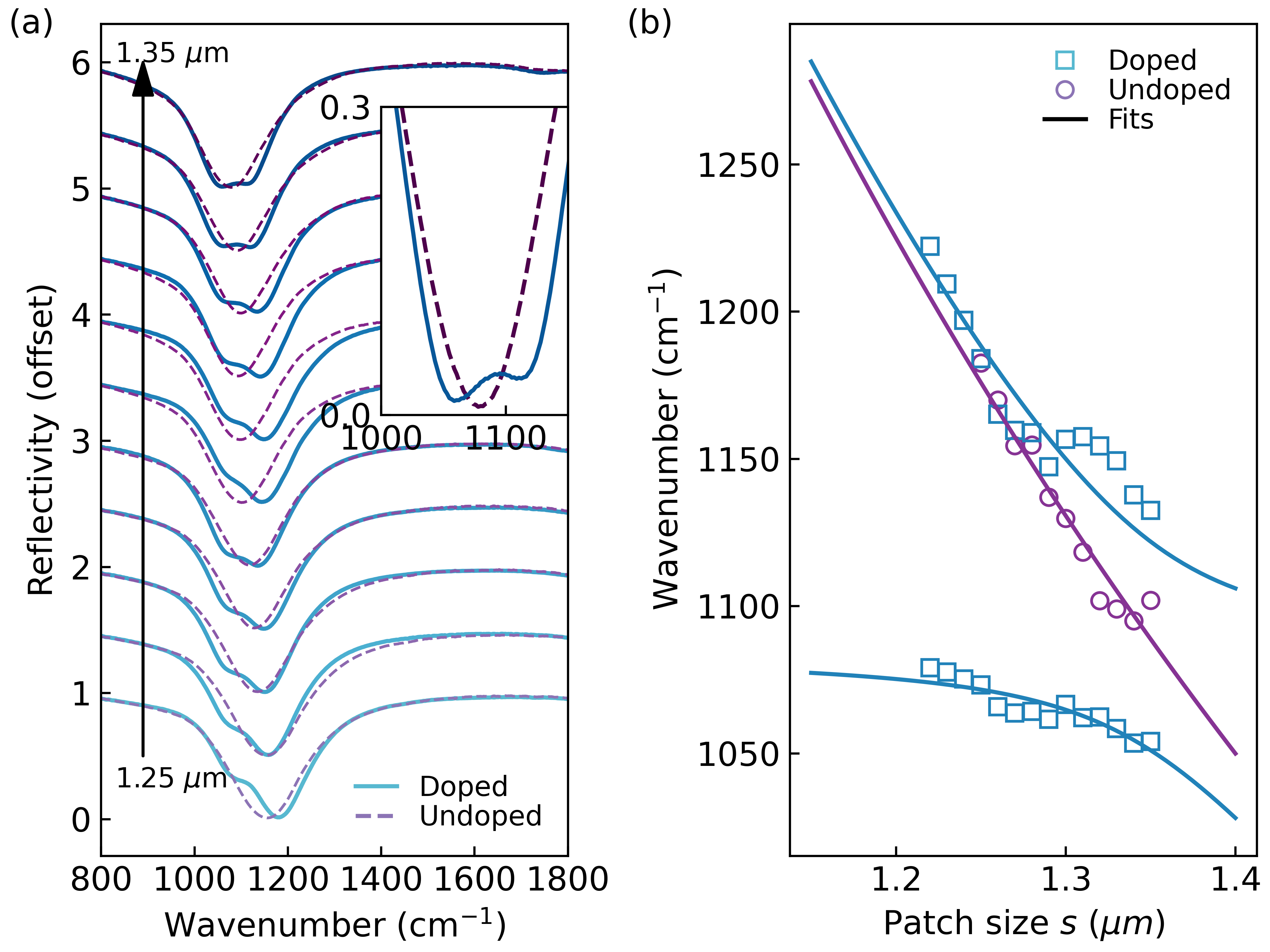}
\caption{\label{fig:disprel}
(a)~Reflectivity of the bare cavity arrays (purple dashed line) and of the doped cavity arrays (blue solid lines) 
as a function of the cavity size. The spectra are offset by 0.5. 
Inset: zoom on the $s=1.35$~{\textmu}m\ spectrum around the polariton splitting.
(b)~Cavity and polaritons dispersion relation as a function of the patch size. 
The symbols are the experimental data (circles: undoped cavities, squares: doped cavities) 
extracted from Lorentzian fits of the spectra. 
The solid lines are fit to eq.~\eqref{eq:cavdisprel} and eq.~\eqref{eq:poldisprel} respectively.}
\end{figure}

The samples are initially characterized in a Fourier-transform infrared (FTIR) spectrometer equipped with a Cassegrain microscope objective, 
to measure the reflectivity spectrum of each cavity array. 
The results are shown in Fig.~\ref{fig:disprel}(a) for both the undoped QWs sample (purple dashed lines) and 
the doped QWs sample (blue solid line). 
The undoped sample exhibits a single absorption dip, corresponding to ohmic dissipation in the cavity.
The absorption frequency $\omega_c$ decreases with increasing patch size $s$ following the usual relation:
\begin{equation}
\omega_c = \frac{\pi c}{n_{eff} s} 
\label{eq:cavdisprel}
\end{equation}
where $n_{eff}$ is an effective refractive index. 
On the contrary, the reflectivity spectra of the doped QWs sample exhibit two absorption dips, on each side 
of the bare cavity absorption, as highlighted in the inset of 
Fig.~\ref{fig:disprel}. 
We fit the datasets with Lorentzian absorption lines and report the extracted peak frequencies 
as a function of patch size $s$ in Fig.~\ref{fig:disprel}(b) (open symbols). 
The two absorption dips in the doped QWs sample exhibit an anti-crossing behavior, characteristic of the strong light-matter coupling regime. 
The low (resp. high) frequency branch corresponds to the lower (resp. upper) polariton. 
First, the dispersion relation of the (undoped) cavity is fitted (purple solid line) according to eq.~\eqref{eq:cavdisprel}, 
providing $n_{eff} = 3.5$, slightly larger than $n_{GaAs}\approx 3.3$ at these frequencies. This stems from the strong confinement of the electric field between the two metallic plates, together with field leakage and reflection phase at the edges of the cavity. 
Then, the dispersion relation of the polariton branches is fitted using the following secular equation\cite{todorov_ultrastrong_2010}:
\begin{equation}
\left( \omega^2 - \omega_{isb}^2 \right) \left(\omega^2 - \omega_c^2 \right) = 4\Omega_R^2 \omega_c^2
\label{eq:poldisprel}
\end{equation}
where $\omega_{isb}$ is the intersubband transition frequency and $\omega_c$ is the cavity frequency. 
In the limit of very low excitation, we can replace $\Delta n \approx n_s$ in the expression of the Rabi frequency.

The slight discrepancy between the secular equation \eqref{eq:poldisprel} and the data can have several causes, e.g. differences in the fabrication of the two samples resulting in actual patch sizes $s$ slightly different between the two samples.
To correct for this effect, we assign the cavities sizes by comparing the frequency of their higher order modes (unperturbed by the ISB transition).
From the fit we extract a Rabi splitting of $2\Omega_R = 70$~cm$^{-1}$, which is slightly 
below the minimal splitting of 78~cm$^{-1}$ measured in the $s = 1.35$~{\textmu}m\ cavity case. 
These values should be compared with the resonance frequency and the cavity/ISB transition linewidths 
(resp. $\gamma^{tot}_c$ and $\gamma^{tot}_{isb}$), yielding  
$\frac{\Omega_R}{\omega_{isb}} = 0.04$ and $\frac{2\Omega_R}{\gamma^{tot}_{c,isb}} = 0.65$. 
The Rabi frequency is only a fraction of the transition frequency, 
but the Rabi splitting is of the order of the dissipation rates of the system. 
This places the system \textit{de facto} at the onset of the strong light-matter coupling regime, 
as we can still resolve the polariton splitting.

\begin{figure}
\centering
\includegraphics[width=\columnwidth]{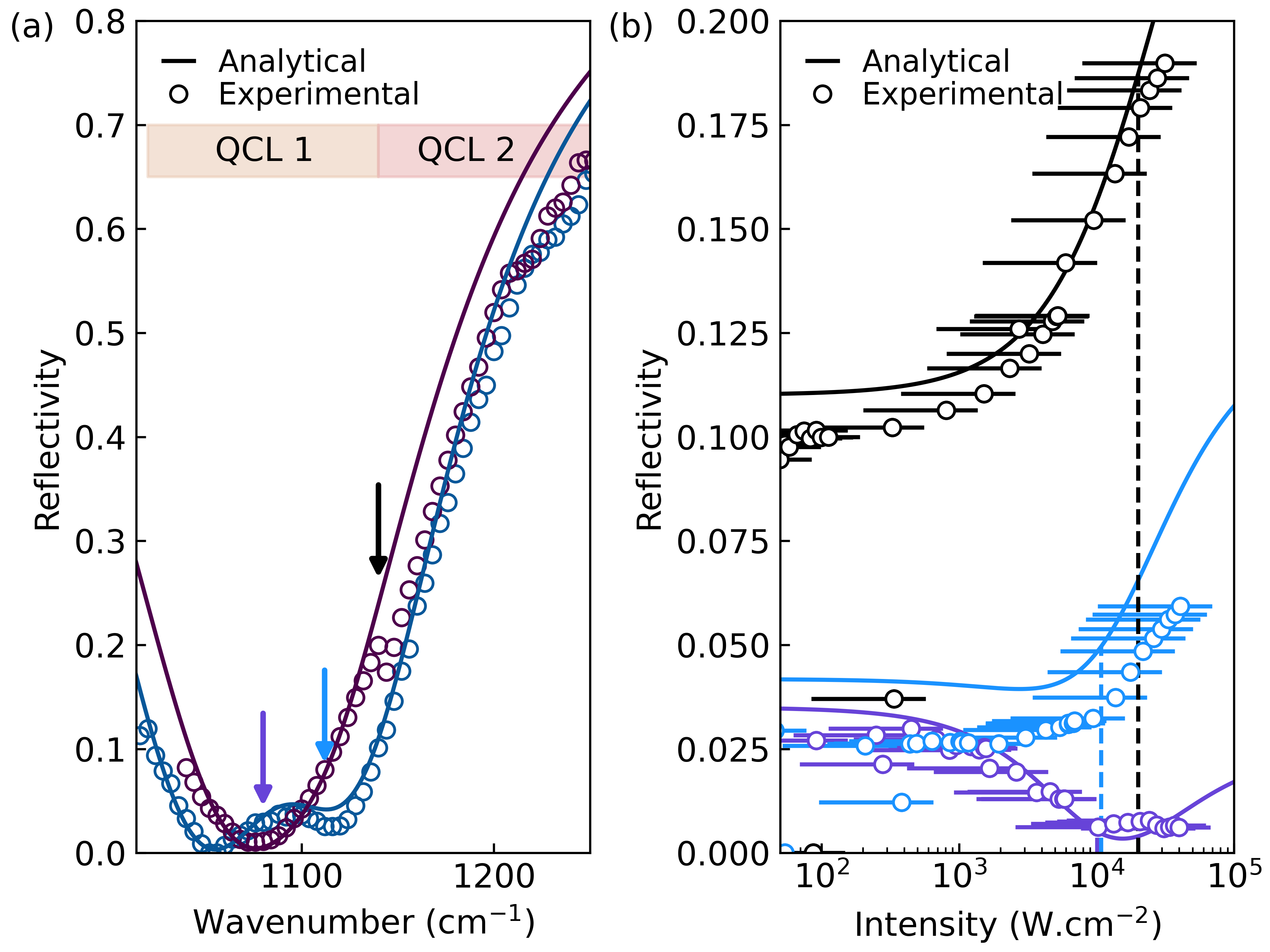}
\caption{\label{fig:saturation}
(a)~Reflectivity spectra of the doped $s=1.35$~{\textmu}m\ cavity array under low (purple circles) and high (blue circles) intensity excitation. 
The shaded areas indicate the tuning ranges of the QCL chips. 
(b)~Nonlinear reflectivity as a function of intensity for three different wavelengths (arrows in (a)). 
Open symbols correspond to experimental data, and solid lines to the CMT prediction. The vertical dashed lines indicate the saturation condition. The horizontal error bars represent uncertainty on the evaluation of the spot size.}
\end{figure}

We then measure the non-linear reflectivity of the doped QWs array closest to resonance ($s = 1.35$~{\textmu}m) 
as a function of the excitation intensity. 
The reflectivity is probed with a home-built microscope setup using a commercial tunable QCL (Daylight MIRCAT) as a source 
and a liquid-nitrogen cooled mercury cadmium telluride detector\cite{pirotta_fast_2021}. 
The QCL emission is amplitude modulated at 10~kHz with pulse widths of 500~ns (0.5\% duty cycle), and lock-in detection is employed. 
The incident intensity can be tuned through a half-wave plate and a polarizer, 
and with the laser injection current (See the Suplementary Material for a complete description).
The reflectivity of the patch cavity array is normalized against that of a smooth Au surface. 
Fig.~\ref{fig:saturation}(a) reports the reflectivity spectra of the sample at low intensity (blue symbols) 
as well as at the maximum intensity available from the QCL (purple symbols). 
For low impinging intensity, we retrieve the two absorption dips (inset of Fig.~\ref{fig:disprel}(a)) 
with two polariton states. On the contrary, the high intensity spectrum displays only one absorption dip, centered between the two polariton states. 

We then select three frequencies from the spectrum in Fig.~\ref{fig:saturation}(a): 
The cavity central frequency (1080~cm$^{-1}$, purple arrow), 
the upper polariton frequency (1112~cm$^{-1}$, light blue arrow) and the 
frequency at which the reflectivity change is the largest (1140~cm$^{-1}$, black arrow). 
Note that the latter frequency accidentally lies at the crossover between two QCL chips, hence the small jump in the high intensity reflectivity spectrum. 
With the laser tuned at these three frequencies we continuously scan the incident intensity. 
The value of the exciting power (in mW) is measured using a thermoelectric detector, and the spot size (20~{\textmu}m~radius at 1/e$^2$) is 
determined with a knife-edge measurement to infer the excitation intensity (in W.cm$^{-2}$). 
The results are presented in Fig.~\ref{fig:saturation}(b) (open symbols). 
Pumping within the polariton gap at the cavity frequency leads to a decrease in reflectivity as the transition saturates, 
and the response of the system converges to the one of the bare cavity. 
On the contrary, pumping at the upper polariton frequency or on the high frequency edge of the spectrum leads to a 
gradual increase in reflectivity. 
%
%


We use temporal coupled mode theory (CMT) to obtain an analytical expression relating the sample reflectivity (and the population difference $\Delta n$) to the incident laser intensity.
The ISB transition and the cavity mode are modeled as oscillators with parameters 
$(\omega_i, \gamma_i, \Gamma_i)$ representing respectively their oscillation frequency, non-radiative, and radiative dampings. 
In the present case, the radiative coupling of the ISB transition to free-space radiation ($\Gamma_{isb} = 0$) can be neglected \cite{alpeggiani_semiclassical_2014, jeannin_absorption_2020}. 
The three equations describing the evolution of the amplitude $a_i$ of each oscillator, 
as well as the in- and out-coupling equation of the exciting field $s^{+}$ and $s^{-}$ are thus\cite{jeannin_unified_2021}:

\begin{align}
    \frac{\mathrm{d}a_{isb}}{\mathrm{d}t} =& (i\omega_{isb}-\gamma_{isb})a_{isb}
        + i\Omega_{R} a_{c}  \\
    \frac{\mathrm{d}a_{c}}{\mathrm{d}t} = & (i\omega_{c}-\gamma_{nr}-\Gamma_{r})a_{c} 
         + i\Omega_{R} a_{isb} + \sqrt{2\Gamma_{r}}s^+ \\
   s^- = & -s^+ + \sqrt{2\Gamma_{r}} a_{c}
\end{align} 

that can be solved in the harmonic regime ($s^{+}=e^{i\omega t}$) to express 
analytically the absorption $\mathcal{A}_{isb} = 2\gamma_{isb}\left|\frac{a_{isb}}{s^+}\right|^2$ 
solely due to the ISB transition, and the reflectivity $R = \left| \frac{s^-}{s^+} \right|^2$. 
Assuming (without loss of generality) that the cavity is at resonance 
with the ISB transition, \textit{i.e.} $\omega_c = \omega_{isb} = \omega_0$, we have:

\begin{widetext}
\begin{align}
\mathcal{A}_{isb}(\Delta n) = & \frac{ 4\gamma_{isb} \Gamma_r\Omega_{R}^2}
			{\left[(\omega-\omega_0)^2-\Omega_{R}^2
				- \gamma_{isb}(\gamma_{nr}+\Gamma_r) \right]^2
			+ \left[(\gamma_{nr}+\Gamma_r+\gamma_{isb})(\omega-\omega_0) \right]^2} \label{eq:Aisb}\\
R(\Delta n) = & \frac{
			\left[(\omega-\omega_0)^2-\Omega_{R}^2
				+\gamma_{isb}(\Gamma_r-\gamma_{nr})\right]^2
			+\left[(\Gamma_r-\gamma_{nr}-\gamma_{isb})(\omega-\omega_0)\right]^2}
		{\left[(\omega-\omega_0)^2-\Omega_{R}^2
				- \gamma_{isb}(\gamma_{nr}+\Gamma_r) \right]^2
			+ \left[ (\gamma_{nr}+\Gamma_r+\gamma_{isb})(\omega-\omega_0) \right]^2} \label{eq:R}
\end{align}
\end{widetext}
These expressions directly depend on the population difference $\Delta n$ through the Rabi frequency \eqref{eq:OmegaR}. 
We then link the excited state excited population of the ISB transition $n_2$ in steady-state condition under an incident intensity $I$ using:
\begin{equation}
n_2 = \frac{I}{N_{qw} \hbar \omega} \tau_{12} \mathcal{A}_{isb}(\Delta n) \label{eq:n2vsI}
\end{equation} 
where $\tau_{12}$ is the upper state lifetime (of the order of the ps) and $N_{qw}$ is the number of QWs. 
Inserting eq.~\eqref{eq:Aisb} in eq.~\eqref{eq:n2vsI} leads - after inversion - to the following intensity-dependent excited state population\cite{baas_optical_2004}:
\begin{widetext}
\begin{equation}
I = \frac{\hbar \omega N_{qw}}{4 \tau_{12} }
		\frac{n_2 \left[ \left[ \gamma_{isb}(\gamma_{nr} + \Gamma_r) 
		- \left( (\omega-\omega_0)^2 -a(n_s - 2n_2) \right) \right]^2 
		+ (\omega - \omega_0)^2 (\gamma_{isb}+\gamma_{nr}+\Gamma_r )^2\right]}
		{\gamma_{isb}\Gamma_r a(n_s - 2n_2)} \\
\label{eq:Ivsn2}
\end{equation}
\end{widetext}
where the coefficient $a$ has been defined in eq. \eqref{eq:OmegaR}. 
Importantly, using \eqref{eq:R},
this result allows us to establish a correspondence between the incident intensity 
and the measured non-linear reflectivity variation through the excited state population. 

All the experimental results (reflectivity spectra and non-linear reflectivity curves) are fitted by combining eq.~\eqref{eq:R} and eq.~\eqref{eq:Ivsn2} and using a single set of parameters. The results are presented as solid lines in Fig.~\ref{fig:saturation}. 
The total quality factor $Q^{tot}_c = \frac{\omega_c}{\Gamma_r+\gamma_{nr}}=6$ and the electronic doping  $n_s=2 \cdot 10^{11}$cm$^{-2}$ are obtained from the experiment: the former one from the experimental total cavity linewidth of the undoped sample (Fig.~\ref{fig:disprel}(a)), and the latter one from the experimental value of the Rabi frequency using eq.~\eqref{eq:OmegaR}.
The only free parameters are the ratio between the radiative and non-radiative cavity dissipation rates, 
and the ISB linewidth $\gamma_{isb}$. 
The fit yields $Q_r = 9.5 \pm 0.3$ and $Q_{nr} = 13 \pm 0.1$, and $Q_{isb}=13 \pm 0.7$. The analytical results are presented in solid lines in Fig.~\ref{fig:saturation}.

The calculations quantitatively reproduce all features of the experimental data. 
We observer the collapse of the polariton states from the low intensity 
to the high intensity reflectivity spectra. 
We also observe the decrease of the non-linear reflectivity as a function of 
intensity when pumping in the gap, and the increase of the non-linear reflectivity when pumping 
at the polariton frequency or on the high frequency edge. 
In particular eq.~\eqref{eq:Ivsn2} permits to deduce the excited state population 
from the non-linear reflectivity curves, an information that cannot be extracted from the sole measurements. 
This allows us to mark the saturation condition \eqref{eq:saturation_condition}, 
shown in vertical dashed lines in Fig.~\ref{fig:saturation}(b). Depending on the pump frequency, saturation 
is achieved for intensities of 10~kW.cm$^{-2}$ to 20~kW.cm$^{-2}$. 

We now examine the limitations of the model and its agreement to experimental data. 
A first approximation is made when considering that the ISB transition frequency does not depend on the 
saturation of the transition. In reality collective effects in confined electron gas renormalize the transition energy 
through the depolarization shift, according to the following formula: 
\begin{equation}
\omega_{isb}^2\left( \Delta n \right) = \omega_{isb,0}^2 + \frac{e^2}{4 \varepsilon \varepsilon_0 m^{\ast} L_{qw}} \Delta n 
\end{equation}
where the square root of last operand on the right-hand-side is known as the plasma frequency.
In mid-IR delta-doped QWs, and at such moderates dopings, the depolarization shift is small~\cite{cominotti_theory_2023}.
The validity of this approximation is gauged {\it a posteriori} by the quantitative agreement between 
theory and experimental results. This effect should be considered at larger doping densities, or in the case of THz ISB transition where the depolarization shift is generally a larger fraction of the transition energy \cite{zaluzny_influence_1993, craig_undressing_1996}.

On the experimental side, we identify two main limitations. The first one is the non-homogeneous excitation 
of the patch cavity array by the (Gaussian) laser beam: all the cavities probed by the beam actually experience different excitation intensities 
and thus exhibit different reflectivity spectra, possibly resulting in a broadening of the response. 
The second limitation is the precise determination of the pump spot size. Given the moderate 
output power of the QCL, a strong focusing is needed to reach high enough intensities. Even when performed with great care, 
knife-edge measurements of the focal spot size are difficult and can lead to variations in the intensity inferred from the power measurement, 
as shown by the error bars in Fig.~\ref{fig:saturation}(b). 

Finally, since the sample is excited using relatively long laser pulses, it is important 
to estimate the importance of thermal effects in the structure and in particular to 
estimate whether the observed effect could simply be a manifestation of heating. 
We report in Supplementary Material an estimation of thermal effects, showing that 
the nonlinear reflectivity change reported here is indeed due to absorption saturation. 

In conclusion, we have demonstrated absorption saturation of an ISB transition at a record low 
intensity (10-20~kW.cm$^{-2}$) by a judicious engineering of the light-matter coupling, 
making it compatible with mid-IR semiconductor lasers. 
The measured non-linear reflectivity properties have validated the analytical model of saturation in weak and strong-coupling regime that we had developed in Ref.~\cite{jeannin_unified_2021}. 
This result constitutes a first step towards semiconductor saturable asborber mirrors (SESAM) able to cover the mid-IR spectral range. 
To further improve this system, future work will focus on improving the main figures of merit of the device: saturation intensity, reflectivity contrast, and non-saturable losses. 
Furthermore, given the fast decoherence times of the order of hundreds of fs \cite{kaindl_homogeneous_2001}, 
and the fast population relaxation times of the order of a few ps \cite{seilmeier_intersubband_1987, lutgen_nonlinear_1996, lutgen_nonequilibrium_1996, vodopyanov_intersubband_1997, mann_ultrafast_2021} of ISB transitions confirmed in recent demonstration of final state stimulation of ISB polaritons \cite{knorr_intersubband_2022}, 
we expect fast saturation and recovery dynamics on the order of few ps\cite{raab_ultrafast_2019}. 
We conclude by emphasizing that this approach is in principle intrinsically scalable to the full mid-IR range.
ISB transitions in III-V semiconductors are well mastered, and can cover the entire mid-infrared. And the cavity electrodynamics concepts enabling the observed ultra-low intensity saturation are also fully scalable to other wavelengths. 
We therefore believe that this experimental demonstration represents the first step towards a robust technology for mid-infrared SESAMs. 
It could have impact for mid-IR lasers, judging from the success SESAMs had at shorter near-infrared wavelengths~\cite{keller_solidstate_1992, keller_semiconductor_1996, keller_recent_2003}. 

\section*{Supplementary Material}
See supplementary material below for a complete description of the non-linear reflectivity setup and an estimation of thermal effects in the structure. 

\section*{Data Availability Statement}
The data that support the findings of this study are available from the corresponding author upon reasonable request.

\begin{acknowledgments}
This work was partially supported by the European Union Future and Emerging Technologies (FET) Grant No. 737017 (MIR-BOSE), and by the French National Research Agency: project SOLID (ANR-19-CE24-0003)
and IRENA (ANR-17-CE24-0016).
This work was done within the C2N micro nanotechnologies platforms and partly supported by the RENATECH network and the General Council of Essonne. 
M.M. acknowledges support from the Marie Skłodowska Curie Action, Grant Agreement No. 748071. 
We acknowledge the technical help of the C2N cleanroom staff. 
We thank Iacopo Carusotto, Ammar Hideur and Adel Bousseksou for scientific discussions, Cristiano Ciuti for granting us access to the Laboratoire Matériaux et Phénomènes Quantiques cleanroom (CNRS UMR 7162, Université Paris-Cité), and Pascal Filloux for assistance in the ICP-RIE etching of the structures.
\end{acknowledgments}

\section{Supplementary - Experimental setup}

\begin{figure}[!ht]
\centering
\includegraphics[width=\columnwidth]{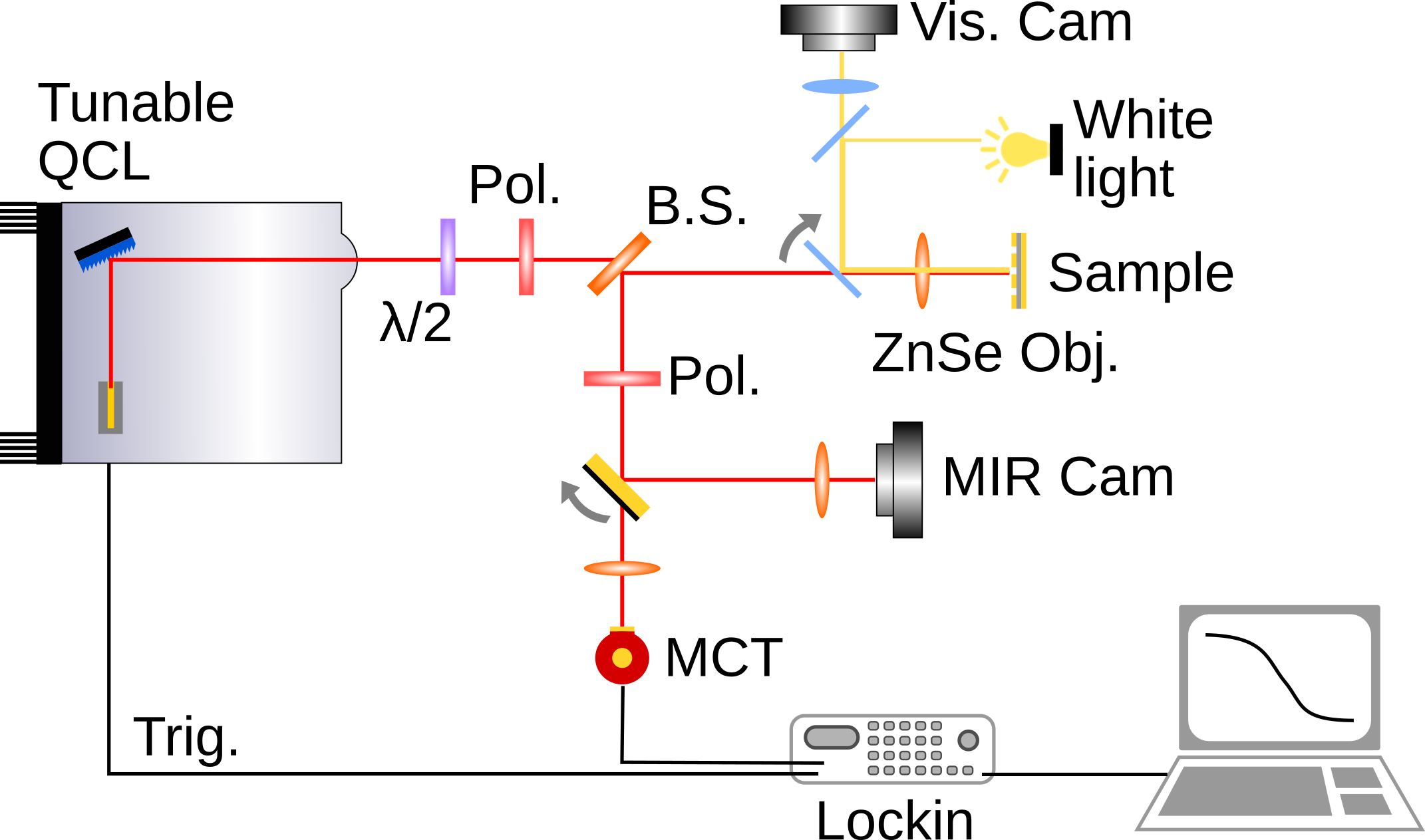}
\caption{\label{fig:setup}
Experimental setup used for the measurement of non-linear reflectivity. 
$\lambda$/2: halfwave plate. Pol.: KRS5 wire grid polarizer. B.S.: 3mm thick 90/10 ZnSe beam splitter. MCT: Mercury Cadmium Telluride detector.
}
\end{figure}

The experimental setup dedicated to the non-linear reflectivity measurement is presented in Fig.~\ref{fig:setup}. 
The tunable QCL is a MIRCAT from Daylight solutions equiped with two chips emitting from 
1024~cm$^{-1}$ to 1140~cm$^{-1}$ (chip 1) and from 1040~cm$^{-1}$ to 1320~cm$^{-1}$ (chip 2), 
in an external cavity configuration with a grating as output coupler. 
The intensity impingin on the sample is controlled by varying either the driving current of the QCL, or by using a combination of halfwave plate and KRS5 wire grid polarizer. 
Light is focused on the sample trough an anti-reflection coated ZnSe objective (Innovation Photonics) with 12~mm focal length. 
The imaging of the laser spot on the sample is done using a mid-IR microbolometric camera (FLIR Systems). The MCT is a J15D19 model from Judsone Teledyne. 
The QCL is driven in pulsed mode with a repetition rate of 10~kHz and pulse widths of 500~ns, and the measurement is performed using a lockin detection technique.

\section{Supplementary - Estimation of thermal effects}

Since the structure is pumped with a moderately intense laser beam 
and relatively long pulses, it is important to consider thermal effects. 
While a comprehensive and detailed thermal analysis of the system is beyond 
the scope of this work, we provide here and estimate 
showing that the observed behavior cannot originate from thermal excitation 
of the carriers. 

\begin{figure}[!ht]
\centering
\includegraphics[width=\columnwidth]{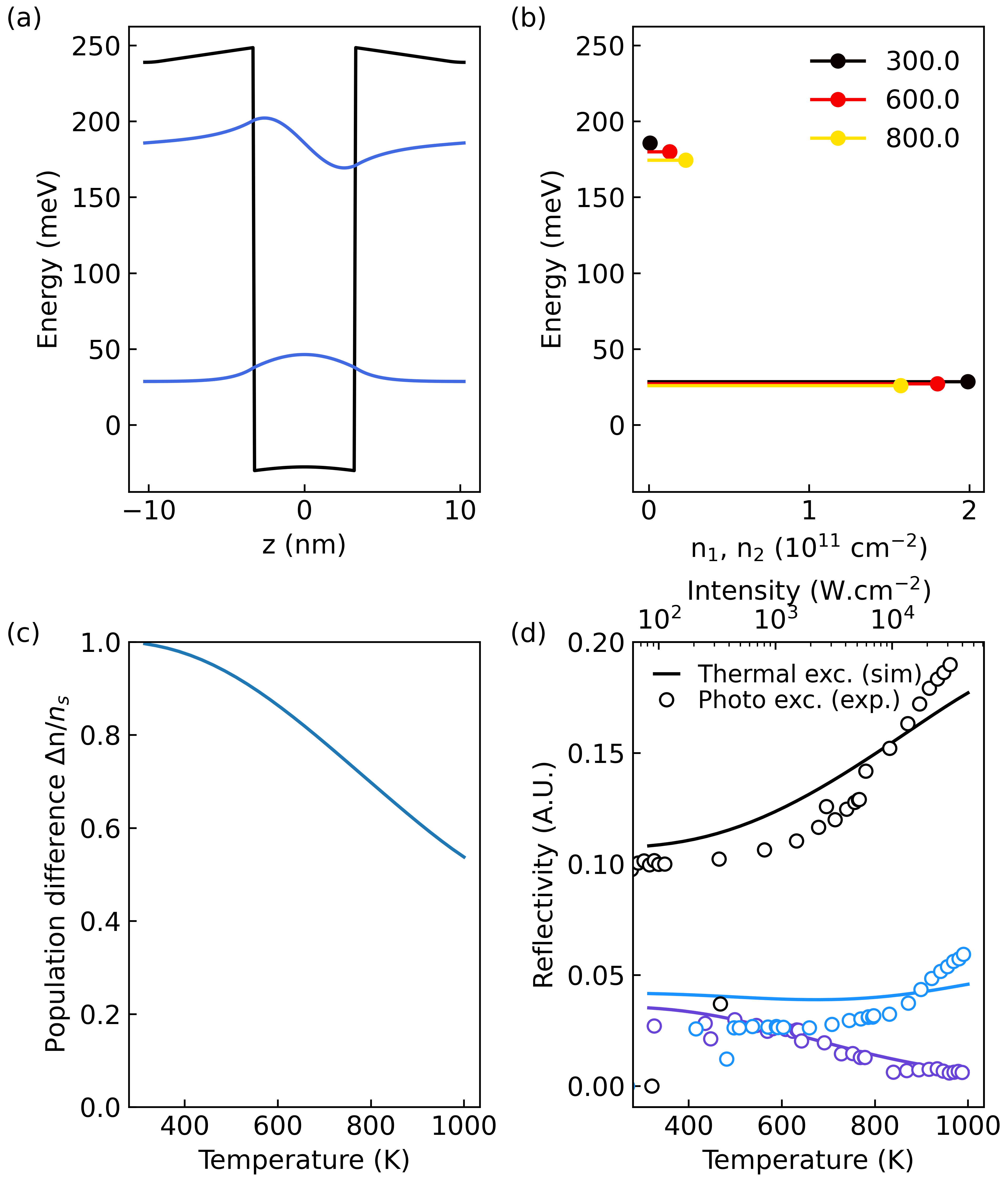}
\caption{\label{fig:temperature}
(a) Band structure of a single QW with the two relevant conduction band states. 
(b) Occupation (carrier densities in $10^{11}$~cm$^{-2}$) of each 
level for three selected temperatures. 
(c) Normalized population difference as a function 
of the temperature. 
(d) Reflectivity as a function of the temperature obtained 
from the CMT model (main text) using the thermal variation of the population in (c) 
as an input.
}
\end{figure}

Using a commercial self-consistent Schrödinger-Poisson solver (NextNano++), we 
simulate the structure for different temperatures from 300~K to 1000~K. 
From the simulation, we can extract the occupation of the first two energy levels in 
the QW, and thus the population difference as a function of the temperature. 
All these results are reported in Fig.~\ref{fig:temperature}. 
Using the same parameters as the main text for the frequencies, dampings and doping, 
we plug the simulated variation of $\Delta n(T)$ (Fig.~\ref{fig:temperature}(c)) 
in the CMT model and obtain the expected reflectivity change as a function of the temperature (Fig.~\ref{fig:temperature}(d), bottom $x$-axis). 
The latter are compared to the experimentally observed variation of the reflectivity as 
a function of the impiging intensity (top $x$-axis). 
One can see that to observe the same level of saturation, the sample temperature should 
reach above 800~K, which should result in alloying of GaAs and Ti/Au layers and destruction of the sample. 
We thus estimate that the observed non-linear reflectivity does not originate from 
thermal effect, and is indeed only based on optical absorption saturation.

\bibliography{Biblio_ISB_SatExp}

\end{document}